\documentclass[12pt]{article}
\usepackage{amssymb}
\usepackage{color,graphicx}
\usepackage[dvipsnames]{xcolor}
\usepackage{amsmath}
\usepackage{array}
\usepackage{epsfig} 
\usepackage{enumerate}
\usepackage{comment}

\usepackage{mathtools}
\newtagform{red}{\color{red}(}{)}
\usepackage{bm}
\def\eps{{\varepsilon}}

\newcommand{\mk}{ \bm k }
\newcommand{\mv}{ \bm v }
\newcommand{\mx}{ \bm x }

\newcommand{\mf}{ \bm f }

\newcommand{\bzero}{ \bm 0 }

\def\dRM{\mathrm{d}}
\def\eRM{\mathrm{e}}

\newcommand{\SA}{ {\mathcal S} }
\def\eps{\varepsilon}
\def\boldnabla{{\bm \nabla}}

\usepackage{xcolor}
\usepackage{comment}

\allowdisplaybreaks
\allowbreak

\begin{document}

\title{Two-species reaction-diffusion system in the presence of random velocity fluctuations}

\author{ Michal Hnati\v{c}$^{1,2,3}$, Matej Kecer$^1$
  Tom\'{a}\v{s} Lu\v{c}ivjansk\'{y}$^{1}$}

\maketitle              
\mbox{ }\\
$^1$ Institute of Physics, Faculty of Science, P. J. Šafárik University, Park Angelinum 9, 040 01 Košice, Slovakia\\
$^2$ Institute of Experimental Physics, Slovak Academy of Sciences, Watsonova 47, 040 01 Košice, Slovakia\\
$^3$ Joint Institute for Nuclear Research, 141980 Dubna, Russia \\

\begin{abstract}
We study random velocity effects on a two-species reaction-diffusion system consisting
of three reaction processes $\textit{A} +\textit{A} \rightarrow (\emptyset, A),$ $\textit{A} +\textit{B} \rightarrow \textit{A}$.
Using the field-theoretic perturbative renormalization group we analyze this system  in the vicinity of its upper critical dimension $d_c = 2$.
Velocity ensemble is generated by means of stochastic Navier-Stokes equations.
In particular, we investigate the effect of thermal fluctuations on reaction kinetics.  The overall analysis is performed to the one-loop approximation and 
possible macroscopic regimes are identified.
\end{abstract}

\section{Introduction}

Diffusion-limited reactions constitute prominent models in non-linear statistical physics \cite{krapivsky2010}.
Theoretical study of such systems attracted a lot of attention in the past \cite{odor2004,tauber2014}.
A straightforward approach to theoretical analysis of such systems is based on kinetic rate equations, which might be regarded as a simple mean-field-like approximation \cite{tauber2014, tauhowlee2005}.
However, reaction systems are known to exhibit non-trivial behavior, especially
 in low space dimensions  \cite{ovchinnikov1989}, where density fluctuations  
become especially pronounced.
 There the kinetic rate equations approach is not adequate and more sophisticated approaches are called for.
%
%
In this paper, we study a multi-species reaction-diffusion system \cite{rajesh2004, lee2018, lee2020, shapoval2022}, which consists of the following three reaction processes 
%
%
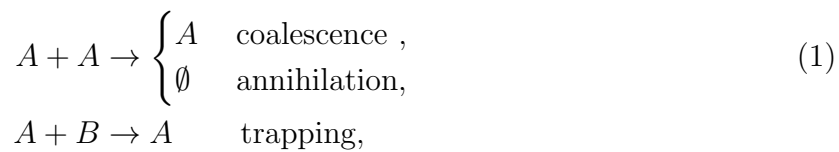
\begin{align}
   A + A &\rightarrow \begin{cases}
   A \hspace{0.5cm} \text{coalescence }, \\
   \emptyset \hspace{0.6cm} \text{annihilation}, \label{reactions}
   \end{cases} \\
   A + B &\rightarrow A \hspace{0.9cm} \text{trapping}, \nonumber
 \end{align}
where coalescence process occurs with probability $p\, (0 \leq p \leq 1)$, and annihilation process with a complementary probability $1-p$. 
The model becomes even more intricate when additional effects are taken into account. Their investigation is especially important, as they naturally arise in many practical circumstances. For instance, the majority of chemical reactions in typical experimental settings occur in some fluid environment. 
Various aspects of such a problem have already been
studied recently \cite{lee2018, lee2020,shapoval2022}. Here, our aim is to
investigate the influence of thermal fluctuations of a surrounding environment on the kinetics of
reaction-scheme \eqref{reactions}. We model the environment as a fluid at
a constant temperature using a well-known approach based on the stochastic
Navier-Stokes equation \cite{forster1977,adzhemyan1999}.

A powerful tool for analyzing the asymptotic behavior
of stochastic systems is provided by the renormalization group (RG) method \cite{vasiliev2004,zinn2002}. It allows us to determine the long-time
and large-scale – or infrared (IR) – asymptotic regimes
of the system and also is a very efficient tool for the calculation
of various universal physical quantities, e.g. critical exponents. 
The aim of this paper is to address the possible IR behavior
of the reaction-diffusion process \eqref{reactions} under the influence of advecting velocity fluctuations and to determine their IR regimes.


The paper is organized as follows. In Sec. \ref{sec:QFT}
we give a field-theoretic formulation of the model and specify the main ingredients of the perturbation theory. Sec. \ref{sec:RG_analysis} is devoted to the analysis of ultraviolet divergences and renormalization of the model in one-loop order of perturbation scheme. The analysis of fixed points (FP) and their regions of stability are discussed in Sec. \ref{sec:RG_functions}. Conclusions are drawn in Sec. \ref{sec:concl}.
\section{Field-theoretic formulation of the model}
\label{sec:QFT}
The field theory for the reaction-diffusion system described by the scheme \eqref{reactions} can be constructed from the master equation by means of Doi-Peliti formalism
\cite{tauhowlee2005,doi1976, peliti1985,lee1994}. For brevity, we omit the derivation as
it can be easily found elsewhere (see e.g., \cite{tauber2014}).
We start our analysis with the field-theoretic action for
the reaction scheme \eqref{reactions} augmented with diffusion processes
\begin{align}
     \SA_{r}[\Psi] &= \psi^\dagger_A(-\partial_t+\nu_0 u_{A0}\partial^2) \psi_A^{~} +  \psi^\dagger_B(-\partial_t+ \nu_0 u_{B0} \partial^2) \psi_B^{~} - \nu_0 u_{A0} \lambda_0 \psi_A^\dagger \psi_A^2 \nonumber\\
     &- \nu_0 u_{A0} \lambda_0 \psi_A^{\dagger 2}\psi_A^2 
    - \lambda_0' Q \nu_0 u_{A0} \psi_B^\dagger \psi_A^{~} \psi_B^{~} - \nu_0 u_{A0} \lambda_0'\psi_A^\dagger \psi_B^\dagger \psi_A^{~} \psi_B^{~}, 
    \label{RDaction}
\end{align}
where $\Psi\equiv\{\psi_A,\psi_A^\dagger,\psi_B, \psi_B^\dagger\}$
are bosonic-like coherent fields arising
 in taking a continuum limit in the Doi-Peliti approach \cite{tauber2014},
 $\partial^2 = \partial_i \partial_i$ denotes Laplace operator in $d$-dimensions and diffusion parameters are expressed through respective Prandtl numbers $u_{A0}$, $u_{B0}$ and viscosity $\nu_0$ (see below Eq. \eqref{NS}). The parameters $\lambda_0, \lambda'_0$ denote reaction constants, and parameter $Q = 1/(2-p)$ is related to the probability of whether annihilation or coagulation process takes place.
In this work,  we employ the RG method, which introduces two different kinds of variables - bare (unrenormalized) quantities and their renormalized counterparts.
 Therefore we denote the former ones with the subscript ``0'', whereas the latter will be written without the subscript ``0''.

Reaction process \eqref{reactions} is an example of a genuine non-equilibrium system and, therefore, we have to specify its initial conditions. 
We choose them in the following form
\begin{equation}
     \SA_{init}[\Psi] = ( a_0 \psi_A^\dagger  +  b_0 \psi_B^\dagger ) \delta(t),
     \label{ic}
\end{equation}
where $a_0,b_0$ are appropriately rescaled initial average 
densities \cite{tauhowlee2005,lee2018}.

In writing actions \eqref{RDaction} and
\eqref{ic} we have employed a condensed notation, in which integrations over space and time variables in the expressions for action functionals are implied. For instance, the first term in the action \eqref{RDaction} 
  corresponds to
\begin{equation}
   - \psi^\dagger_A \partial_t \psi_A = -\int \dRM x \, \psi^\dagger_A (x)
   \partial_t \psi_A (x),   
   \label{eq:shortcut}
\end{equation} 
where we have written coordinates compactly as $x=(t,\mx)$ and integration measure as $\dRM x = \dRM t \dRM^d x$.

The aim of this paper is to study the case where chemical particles are advected within the fluid environment with random fluctuations.
 We introduce advection processes into the formalism by the
  inclusion of convective derivative \cite{Landau_fluid}. 
  This corresponds to the replacement of the time derivative as follows
\begin{equation}
  \partial_t \rightarrow \partial_t + \mv\cdot\boldnabla = \partial_t + v_j \partial_j,   
\end{equation}
where summation over repeated indices is implied in the last term.
Let us stress that the advection for both particle types is considered to be passive, i.e., the velocity field itself is not affected by the particles or reactions processes, respectively.
Corresponding advective terms to the action \eqref{RDaction} take the form 
\begin{equation}
    \SA_{adv}[\Psi,\mv] = -\psi_A^\dagger v_j \partial_j \psi_A^{~} -\psi_B^\dagger v_j \partial_j \psi_B^{~}.
    \label{advection}
\end{equation}
To finalize the model construction we need to specify velocity field $\mv$.
Here, we assume that velocity field $\mv(t, \mx)$ is a random variable with 
zero mean, 
whose dynamics is governed by stochastic Navier-Stokes equation \cite{forster1977,frisch1995}.

\begin{equation}
    \partial_t v_i + (v_j \partial_j) v_i = \nu_0 \partial^2 v_i - \partial_i P + f_i, 
    \label{NS}
\end{equation}
 where $P=P(x)$ is the pressure field, and $f_i=f_i(x)$ denotes $i$-th component of an external random force $\mf$. Following earlier works \cite{forster1977,frisch1995,forster1976} we assume the force $\mf$ is a random Gaussian variable with  zero mean and correlation function
 of the prescribed form
\begin{equation}
    \langle f_i(t,\mx) f_j(0, \bzero) \rangle =  \int 
    \frac{\dRM^d k}{(2\pi)^d} D_{ij}(t,\mk) \eRM^{i\mk \cdot \mx }. 
    \label{F}    
\end{equation}
We consider the case of an incompressible fluid, which implies transversality of the field $\mv$ ($\partial_iv_i = 0$). Using this condition it is possible to express pressure in terms
of velocity field \cite{frisch1995}. This is equivalent to work
in transversal space by
taking the following replacement for velocity field $\mv$ in the momentum representation
\begin{equation}
    v_i(\mk) \rightarrow P_{ij}(\mk) v_j(\mk) ,
\end{equation}
where $P_{ij}(\mk) = \delta_{ij} - k_i k_j /k^2$ with  $(k =|\mk|)$ is transverse projection operator.

%
%
The incompressibility condition implies that the kernel $D_{ij}$ in the momentum representation is proportional to transverse projector $P_{ij}(\mk)$.
 In fact, it can be readily shown that for incompressible medium $D_{ij} \sim \delta_{ij}$ is sufficient. However, we follow the traditional notation
  in previous works and keep $P_{ij}$ in the expression for kernel $D_{ij}$.
%
%
 Using  a specific choice for the momentum dependence of $D_{ij} $ term it is possible to generate  fluctuations of the velocity field near thermal equilibrium.

These considerations finally lead to
\begin{equation}
    D_{ij}(t,\mk) = \delta(t) D_0 k^2 P_{ij}(\mk), \label{Dij}
\end{equation}
where $\delta=\delta(t)$ is Dirac delta function. 
%
%
It can be shown that delta correlations in time of the kernel $D_{ij}$ ensures that the
present model possesses the Galilean symmetry \cite{adzhemyan1999,hhl2019}.
%
%

In hindsight, this particular form \eqref{Dij} is convenient for the application
of RG method, because  both velocity fluctuations and reaction processes of the original reaction-diffusion system become simultaneously marginal in the critical space dimension 
%
%
$d = d_c = 2$.
%
%
The stochastic problem \eqref{NS}-\eqref{Dij} can be recast into a field theory with the doubled set of fields $\Phi = \{ \mv, \mv' \}$ described by the De Dominicis-Janssen action functional \cite{vasiliev2004,adzhemyan1999},
\begin{equation}
      \SA_{v}[\Phi] = \frac{1}{2} v_i' D_{ij} v_j' + v_i' \left( -\partial_t v_i - v_j \partial_j v_i + \nu_0 \partial^2 v_i \right),
      \label{velocityaction}
\end{equation}
where response field $v_i'$ is incompressible, and again condensed notation in the
sense of Eq. \eqref{eq:shortcut} is assumed. Let us note that quadratic
term in the response field $\mv'$ in the action \eqref{velocityaction} actually stands for
%
%
\begin{equation}
    v'_iD_{ij}{v'}_j = \int \dRM x \int \dRM x' 
    v'_i(x)D_{ij}( x - x' ) v'_k(x'), 
\end{equation}
%
%
where $D_{ij}$ corresponds to the inverse Fourier transform of the kernel \eqref{Dij}.

The sum of action functionals 
\eqref{RDaction}, \eqref{ic}, \eqref{advection}, and \eqref{velocityaction}, respectively,  then gives us a final field-theoretic action
\begin{equation}
    \SA = \SA_{r} + \SA_{v} + \SA_{adv} + 
    \SA_{init}.
    \label{bare_action}
\end{equation}
Expectation values of some physical observable $A=A(t,\mx)$ can
be, in principle, calculated 
  as a functional integral \cite{tauhowlee2005,vasiliev2004}
\begin{equation}
    \langle A(t,\mx)  \rangle = \mathcal{N}^{-1} \int \mathcal{D}\Psi \mathcal{D}\Phi\, A(t,\mx) \eRM^{S},
\end{equation}
where $\mathcal{N}$ is a normalization constant.

In what follows we analyze  field-theoretic action \eqref{bare_action} 
using the field-theoretic renormalization group. This technique was employed in the past on similar problems as well 
\cite{tauber2014,park1998,richardson1999,deem1998,hnatic2000, honkonen2002, hnatic2011}.  
We apply it here in a perturbation setting, which is based on expressing
Green functions as a series in coupling constants of a theory.
The perturbation theory of the model is then constructed using well-known Feynman diagrammatic rules \cite{tauber2014, vasiliev2004, zinn2002}. 
The part of the action \eqref{bare_action} quadratic in fields determines the bare propagators, which in frequency-momentum representation take form
\begin{align}    
        \langle \psi_A^{~} \psi_A^\dagger \rangle_0 & =  \frac{1}{-i\omega + \nu_0 u_{A0} k^2},
        &\langle \psi_B^{~} \psi_B^\dagger\rangle_0 & =  \frac{1}{-i\omega + \nu_0 u_{B0} k^2},
        \label{bare_prop1}    
        \\        
        \langle v_i v_j\rangle_0 & = \frac{D_0 k^2 P_{ij}(\mk)}{\omega^2+\nu_0^2 k^4},                 
        &\langle v_iv'_j\rangle_0 & =   \frac{P_{ij}(\mk)}{-i\omega + \nu_0k^2}.    
        \label{bare_prop2}    
\end{align}
The nonlinear terms determine 
interaction vertices with associated vertex factors \cite{vasiliev2004}.
They can be calculated with the help of the formula
\begin{equation}
  V_N(x_1,\ldots,x_N;\varphi) = 
  \frac{\delta^N \SA_{\text{int}}}{\delta\varphi(x_1)\ldots\delta\varphi(x_N)},
  \quad
  \varphi \in\{\psi_A,\psi_A^\dagger,\psi_B,\psi_B^\dagger,\mv,\mv' \},
  \label{eq:ver_factor}
\end{equation}
where $\SA_{\text{int}}$ corresponds to the non-linear terms of the action \eqref{bare_action}.
In a straightforward manner, we get the following bare vertices without an inclusion
of velocity field
%
%
\begin{align}
  V_{ \psi_A^\dagger \psi_A^{~} \psi_A^{~} } & = - 2\lambda_0 \nu_0 u_{A0},
  &V_{ \psi_B^\dagger \psi_B^{~} \psi_A^{~} } & = -\lambda'_0 \nu_0 u_{A0} Q,  \nonumber\\
  V_{ \psi_A^\dagger \psi_A^\dagger \psi_A^{~} \psi_A^{~} }& = -4\lambda_0 \nu_0 u_{A0},
  &V_{ \psi_A^\dagger \psi_B^\dagger \psi_A^{~} \psi_B^{~} }& = -\lambda'_0 \nu_0 u_{A0}.
\end{align}
%
%
On the other hand, there are three additional vertices that include the velocity field 
 \begin{equation}
  V_{ \psi_A^\dagger(\mk) \psi_A^{~} v_j  }  = i k_j, \quad 
   V_{ \psi_B^\dagger(\mk) \psi_B^{~} v_j  }  = i k_j, \quad 
  V_{ v'_i(\mk) v_l v_j }  = i(k_l \delta_{ij} + k_j \delta_{il}). \label{velocity_vertices}
\end{equation}
First, two describe advection processes and the latter vertex is responsible
for interactions between velocity fluctuations.
 Also, we have explicitly written, the momentum of which field enters a given interaction vertex. For instance,
 in expression for the vertex factor 
 $V_{ v'_i(\mk) v_l v_j }$  the momentum $k_j$ is carried by the response
 field $v_i'$ \cite{adzhemyan1999,vasiliev2004}.


\section{Renormalization of the model}
\label{sec:RG_analysis}
The analysis of UV divergences starts with a determination of the canonical dimensions for model parameters. In dynamical models, there are two independent scales that need to be considered \cite{tauber2014, vasiliev2004}. These are frequency and momentum scales (time and length). Then any quantity $F$ is
characterized with both frequency dimension $d_F^\omega$ and a momentum dimension $d_F^k$, respectively. Canonical dimensions are determined from normalization conditions
\begin{equation}
    d_k^k = -d_x^k = 1, \ d_\omega^\omega = - d_t^\omega = 1, \ d_k^\omega = d_\omega^k = 0,
\end{equation}
and the fact that the action functional has to be a dimensionless quantity 
\cite{vasiliev2004}.
The total canonical dimension of any $F$ is then given as $d_F = d_F^k + 2d_F^\omega$ (because of $\partial_t \propto \mathbf{\partial}^2$ proportionality in quadratic part of the action functional). Canonical dimensions of all the fields and parameters of model \eqref{bare_action} are listed in Tab.~\ref{tab:canonical_dimensions}.

\begin{table*}
  \begin{center}
    \begin{tabular}{|c|c|c|c|c|c|c|c|c|c|c|}
      \hline
      $F$ & $\psi_A$, $\psi_B$  & $\psi^\dagger_A$, $\psi^\dagger_B$ & $\mv$ & $\mv'$ & $\lambda_0$, $\lambda'_0$, $g_0$ & $Q$ & $a_0$, $b_0$ & $\nu_0$ & $D_0$ & $u_{A0}$, $u_{B0}$ \\
      \hline
      $d_F^k$ & $d$ & $0$ & $-1$ & $d+1$ & $2-d$ & $0$ & $d$ & $-2$ & $-d-4$ & $0$ \\
      \hline
      $d_F^\omega$ & $0$ & $0$ & $1$ & $-1$ & $0$ & $0$ & $0$ & $1$ & $3$ & $0$ \\
      \hline
      $d_F$ & $d$ & $0$ & $1$ & $d-1$ & $2-d$ & $0$ & $d$ & $0$ & $2-d$ & $0$\\
      \hline
    \end{tabular}
  \end{center}
  \caption{Canonical dimensions of fields and parameters.}
  \label{tab:canonical_dimensions}
\end{table*}

There are altogether five charges (coupling constants) of the theory
\begin{equation}
    g_0 = \frac{D_0}{\nu_0^3}, \ u_{A0}, \ u_{B0}, \ \lambda_0, \ \lambda'_0.
\end{equation}
In the space dimension $d=2$, all of these charges become simultaneously dimensionless and the model becomes logarithmic. Therefore this dimension is identified as an upper critical dimension $d_c$ of the model.
In dimensional regularisation, the UV divergences manifest themselves as poles 
in expansion parameter $\eps = 2-d$, whereas the IR divergences are regulated by the sharp cutoff at $k=m$, which is an analog of the inverse of integral turbulence scale $L = 1/m$. Let us note that the latter divergences do not affect renormalization constants \cite{vasiliev2004}.

Probably the most economical way to renormalize the translationally invariant
model is through the renormalization of its one-particle irreducible (1PI) Green functions.
This is a restricted class of Feynman diagrams  that consists of such diagrams
that remain connected even after one internal line is cut off \cite{tauber2014,vasiliev2004}.
An arbitrary one-particle irreducible (1PI) Green's function will be denoted as
$\Gamma_{\{\varphi\}} = \langle \varphi \ldots \varphi \rangle_{1PI}$, where 
$\varphi \in \Psi \cup \Phi$ denotes an arbitrary field from the full set of fields  of the model \eqref{bare_action}.
Its total canonical dimension is given by a general formula 
\cite{vasiliev2004, zinn2002}
\begin{equation}
    d_\Gamma = d+2 - \sum_\varphi N_\varphi d_\varphi,
\end{equation}
where the sum runs through all the types of fields $\varphi$, $N_\varphi$ denotes the number of times the given field appears in the particular 1PI function and 
$d_\varphi$ is its canonical dimension.
Following the standard approach \cite{vasiliev2004} the task is to identify superficial divergences in 1PI functions and construct renormalized action, in which introduced additional counter-terms ensure the removal of these divergences in the given order of perturbation theory.

The UV divergences, which require further treatment, are identified with those 1PI
 Green functions, which possess a non-negative formal index of divergence $\delta_\Gamma = d_\Gamma|_{\eps = 0}$.
However, for the present case, this statement is to be adjusted based on the following considerations. First, the 1PI functions not involving any of the response functions $\psi_B^\dagger, \psi_A^\dagger, \mv'$ as external fields vanish as they necessarily contain closed cycles of causal propagators \cite{vasiliev2004}.
Since vertex factor $V_{\mv' \mv \mv} $ is proportional to the momentum carried by field $\mv'$ (see the corresponding expression in \eqref{velocity_vertices}), every instance of $\mv'$ appearing as external field lowers the overall index of divergence. Thus the real index of divergence is defined as
\begin{equation}
    \tilde{\delta}_\Gamma =  \delta_\Gamma - N_{\mv'}.
\end{equation}\\
Second, the number of counter-terms is further reduced because of the invariance property of generating functional of model \eqref{bare_action} with respect to Galilean transformations.
This symmetry implies that the function  $\langle \mv' \mv \mv \rangle_{1PI}$ does not diverge (for further discussions on the subject see e.g. \cite{adzhemyan1999, vasiliev2004, hnatic2019}).
Taking these into account along with available diagrammatic elements and transversality of the velocity field 
we can identify the following irreducible functions with superficial UV divergences
%
%
\begin{align}    
        &\langle \mv' \mv' \rangle_{1PI},
        &\langle \psi_A^\dagger \psi_A^{~} \psi_A^{~} \rangle_{1PI},& 
        \nonumber\\
        &\langle \mv' \mv \rangle_{1PI},
        &\langle \psi_B^\dagger \psi_B^{~} \psi_A^{~} \rangle_{1PI},&
        \nonumber\\
        &\langle \psi_A^\dagger \psi_A^{~} \rangle_{1PI},        
        &\langle \psi_A^\dagger \psi_A^\dagger \psi_A^{~} \psi_A^{~}\rangle_{1PI},&    
        \nonumber\\
        &\langle \psi_B^\dagger \psi_B^{~} \rangle_{1PI},
        &\langle \psi_B^\dagger \psi_A^\dagger \psi_B^{~} \psi_A^{~} \rangle_{1PI}&.    
        \label{divergent_1PI}
\end{align}
%
%
All of these have the form that is already present in the bare action functional. This implies that the model is multiplicatively renormalizable. 

The total renormalized action takes the form
%
%
\begin{align}
    S_{R} &=  \psi^\dagger_A \left( -\partial_t+ Z_1 u_A \nu \partial^2 \right) \psi_A^{~} +  \psi^\dagger_B \left( -\partial_t + Z_2 u_B \nu \partial^2 \right) \psi_B^{~} 
    + \frac{ v_i' \mu^\epsilon Z_3 D_{ij} v_j'}{2} 
    \nonumber \\ 
    & + v_i' (-\partial_t  + Z_4 \nu \partial^2) v_i    
    - u_{A} \nu \lambda \mu^\epsilon Z_5 \left[ \psi_A^\dagger  + \psi_A^{\dagger 2}  \right]\psi_A^2 - \lambda' u_{A} \nu \mu^\epsilon Z_6 \left[Q  \psi_A^{~} 
    + \psi_A^\dagger  \psi_A^{~}  \right] 
    \psi_B^\dagger\psi_B^{~}
    \nonumber \\
    & - v_i' (\mv \cdot \boldnabla ) v_i
    - \left[ \psi_A^\dagger (\mv \cdot \boldnabla) \psi_A^{~}  + \psi_B^\dagger 
    ( \mv \cdot \boldnabla ) \psi_B \right] +  
    \delta(t) \left( \psi_A^\dagger \  a_0 + \psi_B^\dagger \ b_0 \right) ,
    \label{renorm_action}
\end{align}
%
%
and was obtained from the bare action \eqref{bare_action} by introducing the following renormalization of fields and parameters of the model
%
%
\begin{align}
    \varphi & \rightarrow Z_{\varphi} \varphi,     
    & u_{A0} & \rightarrow Z_{u_A} u_A,
    & u_{B0} & \rightarrow Z_{u_B} u_B, 
    & \nu_0 & \rightarrow Z_\nu \nu,  \nonumber\\
     g_0 & \rightarrow \mu^\eps Z_g g,
    & \lambda_0 & \rightarrow \mu^\eps Z_\lambda \lambda,
    & \lambda'_0 & \rightarrow \mu^\eps Z_{\lambda'} \lambda'.
\label{renorm_parameters}
\end{align}
%
%
Here, $\varphi \in\{\psi_A,\psi_A^\dagger,\psi_B,\psi_B^\dagger,\mv,\mv' \}$, $\mu$ is an arbitrary momentum scale and 
$Z_F$  denotes the corresponding renormalization constant. 

By direct inspection, we get relations between renormalization constants in the renormalized action \eqref{renorm_action} and  RG constants \eqref{renorm_parameters}
%
%
\begin{align}
    Z_{u_A} &= Z_1 Z_4^{-1}, 
    &Z_{u_B}& = Z_2 Z_4^{-1}, 
    &Z_\lambda& = Z_5 Z_1^{-1}, 
    &Z_\lambda'& = Z_6 Z_1^{-1},  \nonumber\\
    Z_g &= Z_3 Z_4^{-3}, 
    &Z_\nu& = Z_4,     
    &Z_\Psi& = Z_Q = 1. 
\end{align}
%
%

The explicit form of RG constants $Z_1 - Z_6$ is calculated from the one-loop 1PI Feynman diagrams using dimensional regularisation and a minimal subtraction scheme. 
The final expressions read
%
%
\begin{align}
    &Z_1 =  1 -  \frac{ \hat{g} }{4u_A(u_A+1) \eps }, 
    &Z_2& = 1 -  \frac{ \hat{g} }{4u_B(u_B+1) \eps }, 
    &Z_3& =  Z_4 = 1 -  \frac{\hat{g}}{16 \eps }, \nonumber\\
    & Z_5  = 1 + \frac{\hat{\lambda}}{ \eps }, 
    & Z_6 & = 1 +  \frac{\hat{\lambda}' u_A}{(u_A+u_B)\eps},
\end{align}
%
%
where $\hat{F} \equiv F S_d/(2\pi)^d$, 
$S_d=2\pi^{d/2}/\Gamma(d/2)$ is the area of unit $d$-dimensional sphere, and 
 $\Gamma(x)$ is Euler’s gamma function.

\section{RG functions and scaling regimes}
\label{sec:RG_functions}
Once the calculation of RG constants is successfully accomplished, it is possible to analyze the asymptotic behavior of the system.
A fundamental equation that governs the behavior of renormalized Green functions
is expressed with the help of the RG operator, which can be expressed in the present case  as
\begin{equation}
    D_{RG} = \mu \partial_\mu + \sum_e \beta_e \partial_e - \gamma_\nu \nu \partial_\nu,
\end{equation}
where the given sum runs through all charges of the theory $e = \{ g, u_A, u_B, \lambda', \lambda \}$. The coefficient functions are defined as
\begin{equation}
    \beta_e = \mu \frac{\partial e}{\partial \mu}\biggr|_0, \quad \gamma_F = \frac{\partial \ln Z_F}{\partial \ln\mu}\biggr|_0, \label{def_rg_functions}
\end{equation}

for any parameter, $F$ and $|_0$ means that bare parameters are held constant during evaluation.
For the model \eqref{bare_action}, we have altogether five beta functions
%
%
\begin{align}
    \beta_g =& -g(\eps + \gamma_g), 
    &\beta_{u_A}& = -u_A\gamma_{u_A}, 
    &\beta_{u_B}& = -u_B \gamma_{u_B}, \nonumber \\
    \beta_\lambda =& -\lambda(\eps + \gamma_\lambda),
    &\beta_{\lambda'}& = -\lambda' (\eps + \gamma_{\lambda'}),
    \label{beta_functions}
\end{align}
%
%
with corresponding anomalous dimensions (by definition Eq. \eqref{def_rg_functions}) 
%
%
\begin{align}
\gamma_g =& -\frac{\hat{g}}{8}, 
&\gamma_{u_i}& = \hat{g} \biggl( \frac{1}{4 u_i (1+u_i)} - \frac{1}{16}\biggr);i\in\{A,B\}, \nonumber\\
\gamma_\lambda =& -\hat{\lambda}- \frac{\hat{g}}{4u_A(1+u_A)},
&\gamma_{\lambda'}& = -\hat{\lambda'} \frac{u_A}{u_A + u_B} - \frac{\hat{g}}{4u_A(1+u_A)},
\label{gamma_functions}
\end{align}
%
%
where the higher order corrections $\hat{g}^2, \hat{\lambda}^2$ are neglected in the one-loop
approximation.

The long-time asymptotic behavior of the model is governed by the IR stable fixed points (FP) \cite{vasiliev2004,zinn2002} of beta functions.
These are such points 
$e^* = ( g^*, u_A^*, u_B^*, \lambda^*, \lambda'^{*} )$ in coupling constant space that 
satisfy
\begin{equation}
    \beta_g(e^*) = \beta_{u_A}(e^*) = \beta_{u_B}(e^*) = \beta_\lambda(e^*) = \beta_{\lambda'}(e^*) = 0.
\end{equation}
IR stability is determined by the eigenvalues of the matrix of the first derivatives
\begin{equation}
    \Omega_{ij} = \frac{\partial \beta_i}{\partial e_j}\biggl|_{e^*},
    \label{eq:omega_matrix}
\end{equation}
where index $i$ and charge $e_j$ belong to the set $\{ g, u_A, u_B, \lambda', \lambda \}$.
For IR stable regimes the eigenvalues of the matrix \eqref{eq:omega_matrix} have
to have positive real parts. We have found eight FPs, however, only two of them are IR stable (see Tab. \ref{tab:fixed_points}). These are
\begin{enumerate}
    \item Gaussian fixed point (FP1): $g^* = 0$, $u_A^* =$ arbitrary, $u_B^* =$ arbitrary, $\lambda^* = 0$, $\lambda'^* = 0$. IR stable for $\eps<0$.
    \item Thermal fixed point (FP8): $g^* = 8\eps$, $u_A^* = u_B^* =(\sqrt{17} - 1)/2$, $\lambda^* = \eps/2$, $\lambda'^* = \eps$. IR stable for $\eps>0$.
\end{enumerate}
Let us note that in non-trivial (thermal) FP both velocity fluctuations and reaction interactions are simultaneously IR-relevant.
RG predicts also an FP for which only reaction processes are relevant (FP4). 
%
%
However, even though it would have been stable without the velocity field \cite{lee1994}, it can never be truly IR stable in the presence of thermal fluctuations, which are inevitable in practice.
A similar conclusion was obtained in the past for different reaction-diffusion model \cite{hnatic2000}.
%
%
%
%
 On the borderline of the two regimes, i.e. for
 the case $\eps = 0$, couplings of the theory become marginally irrelevant, and logarithmic corrections are expected to appear in expressions for Green's functions. Based on the standard analysis \cite{vasiliev2004} we predict that these 
 corrections will be different from the ones realized if the velocity field was not present (and FP4 would be stable). Therefore the behavior in two-dimensional systems is also expected to be affected by the presence of velocity field fluctuations. The proof of this statement is deferred to future work.
%
%

\begin{table*}[t]
  \begin{center}
 \setlength{\extrarowheight}{2pt}
    \begin{tabular}{|c|c|c|c|c|c|}
    \hline
      Fixed point & $\hat{g}^*$ & $u_A^*, u_B^*$ & $\hat{\lambda}^*$ & $\hat{\lambda}'^*$ & Stability region \\
      \hline
     FP1 & $0$ & arbitrary & $0$ & $0$ & $\eps <0$\\
      \hline
    FP2 & $0$ & arbitrary & $0$ & $\eps (u_A+u_B)/u_A$ & unstable\\
      \hline
    FP3 & $0$ & arbitrary & $\eps$ & $0$ & unstable\\
      \hline
    FP4 & $0$ & arbitrary & $\eps$ & $\eps (u_A+u_B)/u_A$ & unstable\\
      \hline
    FP5& $8\eps$ & $(\sqrt{17} - 1)/2$ & $0$ & $0$ & unstable\\
      \hline
    FP6& $8\eps$ & $(\sqrt{17} - 1)/2$ & $0$ & $\eps$ & unstable\\
      \hline
    FP7& $8\eps$ &$(\sqrt{17} - 1)/2$ & $\eps/2$ & $0$ & unstable\\
      \hline
    FP8& $8\eps$ & $(\sqrt{17} - 1)/2$ & $\eps/2$ & $\eps$ & $\eps > 0$\\
    \hline
    \end{tabular}
  \end{center}
  \caption{RG fixed points and their region of stability.}
  \label{tab:fixed_points}
\end{table*}
{ \section{Conclusion} \label{sec:concl} }
We have investigated the influence of thermal fluctuations on a reaction-diffusion system with reactions $\textit{A} +\textit{A} \rightarrow (\emptyset, A),$ $\textit{A} +\textit{B} \rightarrow \textit{A}$. Using the field-theoretic formulation of the model we
have analyzed possible macroscopic behavior utilizing the renormalization group approach.  
In particular, we have renormalized the model to the one-loop order of the perturbation scheme. The RG analysis revealed the existence of two IR-stable FPs which govern the 
long-time behavior of the system. 
\section*{Conflict of Interest}
 The authors declare that they have no
conflicts of interest.
\section*{Acknowledgment}
The work was supported by VEGA grant No. 1/0535/21 of the Ministry of Education, Science, Research and Sport of the Slovak Republic.

Conflict of Interest: The authors declare that they have no conflicts of interest.

\end{document}